# Pressure-induced superconductivity up to 9 K in the quasi-one-dimensional KMn$_6$Bi$_5$

Z. Y. Liu[1,2=], Q. X. Dong[1,2=], P. T. Yang[1,2=], P. F. Shan[1,2], B. S. Wang[1,2*], J. P. Sun[1,2], Y. Uwatoko[3], G. F. Chen[1,2*], X. L. Dong, Z. X. Zhao, J.-G. Cheng[1,2*]

[1]*Beijing National Laboratory for Condensed Matter Physics and Institute of Physics, Chinese Academy of Sciences, Beijing 100190, China*

[2]*School of Physical Sciences, University of Chinese Academy of Sciences, Beijing 100190, China*

[3]*Institute for Solid State Physics, University of Tokyo, Kashiwa, Chiba 277-8581, Japan*

= These authors contributed equally to this work.

Correspondence should be addressed to B.S.W. (bswang@iphy.ac.cn), G.F.C. (gfchen@iphy.ac.cn) or J.G.C. (jgcheng@iphy.ac.cn)

## Abstract

The Mn-based superconductor is rare owing to the strong magnetic pair-breaking effect. Here we report on the discovery of pressure-induced superconductivity in KMn$_6$Bi$_5$, which becomes the first *ternary* Mn-based superconductor. At ambient pressure, the quasi-one-dimensional KMn$_6$Bi$_5$ is an antiferromagnetic metal with $T_N \approx 75$ K. By measuring resistivity and ac magnetic susceptibility under hydrostatic pressures up to 14.2 GPa in a cubic anvil cell apparatus, we find that its antiferromagnetic transition can be suppressed completely at a critical pressure of $P_c \approx 13$ GPa, around which bulk superconductivity emerges and displays a superconducting dome with the maximal $T_c^{onset} = 9.3$ K achieved at about 14.2 GPa. The close proximity of superconductivity to a magnetic instability in the temperature-pressure phase diagram of KMn$_6$Bi$_5$ and an unusually large $\mu_0 H_{c2}(0) = 18.9$ T exceeding the Pauli limit suggests an unconventional magnetism-mediated paring mechanism. In contrast to the binary MnP, the flexibility of the crystal structure and chemical compositions in the ternary $A$Mn$_6$Bi$_5$ ($A$ = alkali metal) can open a new avenue for finding more Mn-based superconductors.

## Introduction

Unconventional superconductivity beyond the Bardeen-Cooper-Schrieffer (BCS) theory usually emerges on the border of long-range antiferromagnetic order as exemplified by the cuprates [1,2], iron-pnictides [3,4], and heavy-fermion superconductors [5,6]. In these cases, the enhanced magnetic fluctuations are believed to be the binding glue of superconducting Cooper pairs. As such, it is expected to find unconventional superconductors in the vicinity of antiferromagnetic quantum critical point (QCP), which can be achieved by employing external tuning parameters such as the chemical doping and/or applying physical pressure [1-6]. By following this approach, we have discovered the first Mn-based superconductor MnP, by suppressing its helimagnetic order via the application of high pressure [7]. Its superconducting transition temperature $T_c \approx 1$ K is very low and the superconductivity only appears in a narrow pressure range around $P_c \approx 8$ GPa where the helimagnetic order just vanishes. Since MnP is a three-dimensional binary compound, other tuning methods except for applying high pressure are less effective in regulating the magnetism and inducing superconductivity so far. In this regard, it is more attractive to achieve superconductivity in the ternary or complex Mn-based compounds, especially those having low-dimensional structures in recognizing the common features of known high-$T_c$ superconductors.

To this end, we turn our attention to a newly synthesized ternary compound $KMn_6Bi_5$ [8], which adopts an interesting quasi-one-dimensional (Q1D) monoclinic structure. As illustrated in Fig. 1(a), the most prominent structural feature of this compound is the presence of infinite $[Mn_6Bi_5]^-$ columns composed of an outer nanotube of Bi atoms (diameter $\sim 8.7$ Å) surrounding an inner Mn-Mn bonded metallic core. The inner core can be described as a vertice-sharing distorted Mn-centered icosahedra along the column direction, i.e. the $b$-axis. The short interatomic Mn-Mn and Bi-Bi distances in comparable to those in the corresponding metal elements render a metallic conductivity for this compound, and the Q1D character produces a pronounced anisotropy in physical properties. Upon cooling down at ambient pressure, it undergoes an antiferromagnetic transition at $T_N \approx 75$ K, which is manifested as a kink in resistivity $\rho_{//}$ along the rod but an upturn in the $\rho_\perp$ perpendicular to the rod, respectively [8]. Analyses of the paramagnetic susceptibility $\chi(T)$ above $T_N$ in terms of the modified Curie-Weiss law yield relatively small effective moments of Mn, i.e. 1.56 $\mu_B$ for $\chi_{//}$ and 1.37 $\mu_B$ for $\chi_\perp$, respectively, signaling an itinerant nature of the magnetism similar to MnP [7]. First-principles calculations on its sister compound $RbMn_6Bi_5$ have revealed that the density of states at Fermi level are dominated by the Mn-3d electrons and the helical antiferromagnetic structures are stable [9]. The itinerant-electron magnetism with relatively low $T_N$ makes $KMn_6Bi_5$ a potential candidate for exploring unconventional superconductivity near the magnetic QCP. In particular, the enhanced magnetic fluctuations pertinent to the strong Q1D structure are expected to promote stronger superconducting pairing and thus higher $T_c$ in comparison to the three-dimensional MnP [7].

We thus employed a high-pressure approach to explore superconductivity in $KMn_6Bi_5$ single crystal by measuring its electrical transport and ac magnetic susceptibility under various pressures up to 23 GPa. We find that its antiferromagnetic order is first enhanced moderately and then suppressed completely at $P_c \approx 13$ GPa, around which bulk superconductivity emerges and exhibits a dome-like $T_c(P)$ with the maximal $T_c^{onset}$ = 9.3 K achieved at 14.2 GPa. The close proximity of superconductivity to a magnetic instability indicates an unconventional paring mechanism and $KMn_6Bi_5$ becomes the first ternary Mn-based superconductor. The present work can open a new avenue for finding more Mn-based superconductors in the ternary $AMn_6Bi_5$ ($A$ = alkali metal) with flexible crystal structure and different chemical variants.

## Experimental

The $KMn_6Bi_5$ crystals used in the present study were grown with the self-flux method as described elsewhere [8]. The quality of the crystals were checked by single-crystal X-ray diffraction (XRD) at room temperature, and the obtained lattice parameters in the monoclinic structure ($C2/m$) are consistent with previous reports [8]. The chemical composition, K:Mn:Bi = 1:6:5, close to the stoichiometric one is also verified via the energy dispersive X-ray spectroscope (EDX) measurements. Electrical transport and magnetic properties at ambient pressure (AP) were measured with the Quantum Design Physical Property Measurement System (PPMS-9T) and Magnetic Property Measurement System (MPMS-III), respectively. High-pressure electrical resistivity measured along the $b$-axis with standard four-probe configurations up to 14.2 GPa were carried out with a palm-type cubic anvil cell (CAC), which can generate excellent hydrostatic pressures due to the three-axis compression geometry and the adoption of liquid pressure transmitting medium (PTM). AC magnetic susceptibility under pressure was also measured in CAC by mutual induction method at a fixed frequency of 317 Hz with ac magnetic field of few Oe parallel to the *b*-axis. The primary and secondary coils are made of enameled cooper wires of 25 μm in diameter with a total of ~ 40 turns for each coil. The glycerol is employed as the PTM for the high-pressure measurements in CAC, and the pressures in CAC were estimated based on the pre-determined pressure calibration curve at room temperature or the superconducting transition of Pb at low temperatures [10,11]. We also measured resistance under quasi-hydrostatic pressures up to 23 GPa by using a BeCu-type diamond pressure cell (DAC) with 300-μm culets. In this case, soft KBr powder was employed as the solid PTM. All the experiments were carried out in a $^4$He refrigerated cryostat equipped with a 9T superconducting magnet at the Synergic Extreme Condition User Facility (SECUF).

## Results and discussion

Before performing the high-pressure studies, we first characterized the physical properties of $KMn_6Bi_5$ single crystal at AP. Figure 1(b) shows the temperature dependences of resistivity $\rho(T)$ and magnetic susceptibility $\chi(T)$ along the *b*-axis. All results are in excellent agreement with those in the previous report [8], and consistently confirm the occurrence of an antiferromagnetic transition at $T_N \approx 75$ K, manifested as a weak kink in $\rho(T)$, a peak in $d\rho/dT$ and a sudden drop in $\chi(T)$. The residual resistivity ratio ($RRR \equiv \rho_{300K}/\rho_{2K}$) is ~ 17 in this study, comparable to ~ 19 in Ref. 8, further confirming a high quality of the studied $KMn_6Bi_5$ crystals. We also noticed that $\rho(T)$ decreases quickly while $\chi(T)$ exhibits a broad shoulder centered at $T^* \approx 40$ K, which might be correlated with the rearrangement of Mn spins [8].

Figure 2(a) shows the resistance $R(T)$ of $KMn_6Bi_5$ (#1) under various pressures up to 14.2 GPa in the temperature range 1.5-300 K. With increasing pressure up to 4 GPa, the $R(T)$ and its derivative $dR/dT$ retains similar temperature dependence to those at AP, the values of $T_N$ raise slightly to ~ 86 K at 4 GPa. At $P \geq 6.2$ GPa, the downward kink-like feature around $T_N$ in $R(T)$ changes to a weak upward hump-like anomaly, manifested as a corresponding change of $dR/dT$ from a peak to a broad dip as shown in Fig. 2(b) and Fig. S1. Such a change should be ascribed to a pressure-induced subtle modification of magnetic structure in $KMn_6Bi_5$, which can affect the electronic states at Fermi level and the corresponding magnetic scatterings. Upon further increasing pressure, the resistance anomaly at $T_N$ is further weakened and the corresponding $T_N$ is lowered progressively, reaching ~ 25 K at 12.9 GPa, above which no anomaly can be discerned in $R(T)$. This indicates that the long-range antiferromagnetic order of KMn$_6$Bi$_5$ is completely suppressed by pressure.

Accompanied with the gradual suppression of antiferromagnetic transition by pressure, we observed a sudden drop of resistance at low temperatures above 12.5 GPa, indicating the possible occurrence of superconductivity in KMn$_6$Bi$_5$. The enlarged view of the low-temperature $R(T)$ data shown in Fig. 3(a) depicts clearly the evolution of the superconducting transition. As shown, $R(T)$ at 12.5 GPa starts to drop quickly at ~ 4 K but cannot reach zero down to 1.5 K, the lowest temperature in the present study. The superconducting transition moves quickly to higher temperatures with increasing pressure and zero-resistance state is realized at $P \geq 12.9$ GPa. Since the observed transition is relatively broad, here we define the onset and offset temperatures of the superconducting transition, $T_c^{\text{onset}}$ and $T_c^{\text{offset}}$, as the intersection points of two straight lines below and above the transition as illustrated in Fig. 3(a). At 14.2 GPa, $T_c^{\text{onset}}$ and $T_c^{\text{offset}}$ are found to reach about 9.3 and 7.6 K, respectively. It is noted that the increment of $T_c$ with pressure becomes less efficient above 13.5 GPa, leading to a saturation trend of $T_c(P)$ as shown below.

To confirm the bulk nature of the superconducting state, we measured the ac magnetic susceptibility $\chi_{ac}(T)$ of KMn$_6$Bi$_5$ (#2) together with a piece of Pb; the superconducting shielding volume fraction $4\pi\chi_{ac}(T)$ of KMn$_6$Bi$_5$ can be estimated by comparing its diamagnetic signal with that of Pb. As shown in Fig. 3(b), the superconducting transition of Pb is quite sharp and moves down progressively with increasing pressure, elaborating an excellent hydrostatic pressure environment in CAC. In perfect agreement with the $R(T)$ data, the superconducting transition in $\chi_{ac}(T)$ marked by $T_c^{\chi}$ starts to appear around 2.2 K at 12.5 GPa and quickly increases up to ~6 K at 13.8 GPa; accordingly, the $4\pi\chi_V$(1.5 K) increases monotonously with pressure from nearly zero at 12.5 GPa to over 90% at 13.8 GPa. From these above high-pressure transport and magnetic susceptibility measurements, we can conclude that bulk superconductivity emerges in KMn$_6$Bi$_5$ around $P_c \approx 13$ GPa, where the long-range antiferromagnetic transition almost vanishes completely.

To complete the evolution of superconducting state under higher pressure, we measured $R(T)$ of KMn$_6$Bi$_5$ (#3) up to 23 GPa by using a DAC. As shown in Fig.S2, it enters into the superconducting state with a pronounced drop in resistance as marked by $T_c^{\text{onset}}$, but zero resistance cannot be reached due to the non-hydrostatic pressure conditions in DAC. $T_c^{onset}$ increases monotonously from ~ 4.95 K at 12.1 Gpa to a maximum of ~7.0 K at 14.9 Gpa, and then decreases to ~ 3.0 K at 23.1 Gpa. These results suggest the presence of a superconducting dome in KMn$_6$Bi$_5$. It is noted that the maximal $T_c^{onset}$ in DAC is lower than that in CAC, which should be ascribed to the different pressure environments of these two techniques.

To further characterize the superconducting state of KMn$_6$Bi$_5$, we measured the low-temperature $R(T)$ under different magnetic fields at each pressure. All the $R(T)$ data are given in Fig. S3, and the normalized data at 14.2 GPa are shown in Fig. 3(c) as a representative. The superconducting transition shifts to lower temperatures gradually with increasing magnetic fields as expected. However, the application of an 8 T field cannot eliminate the superconducting transition at 14.2 GPa, implying a high upper critical field $\mu_0 H_{c2}$. To quantify the evolution of $\mu_0 H_{c2}$, here we used the criteria of 50% $R_n$ for $T_c$ and plotted the temperature dependences of $\mu_0 H_{c2}(T)$ in Fig.3(d). The experimental data can be well described by the Ginzburg-Landau (GL) equation, $\mu_0 H_{c2}(T) = \mu_0 H_{c2}(0)[1-(T/T_c)^2]/[1+(T/T_c)^2]$ [12]. The best fits shown by the dash lines in Fig. 3(d) yield the zero-temperature $\mu_0 H_{c2}(0)$ values for each pressure. Accordingly, the coherent lengths $\xi(0)$ at each pressure can be obtained based on the equation $\mu_0 H_{c2}(0) = \Phi_0/2\pi\xi(0)^2$, where $\Phi_0 = hc/2e$ is the magnetic flux quantum. The obtained values of $\mu_0 H_{c2}(0)$ and $\xi(0)$ under different pressures are collected in the Table S1. As can be seen, with increasing pressure from 12.9 to 14.2 GPa, the $\mu_0 H_{c2}(0)$ increases dramatically from 7.42 T to 18.95 T, while the $\xi(0)$ decreases from 66.6 to 41.7 Å. It is noteworthy that the $\mu_0 H_{c2}(0)$ at 14.2 GPa exceeds the Pauli limit of $\mu_0 H_p = 1.84T_c = 15.2$ T [13], implying a possible unconventional pairing mechanism.

To ensure the reproducibility of the obtained results, we performed additional high-pressure resistance measurements on KMn$_6$Bi$_5$ (#4) in CAC up to 13.3 GPa. As shown in Figs. S4-S5, both $R(T)$ and the $T_N$ show similar pressure dependences to those of KMn$_6$Bi$_5$ (#3); $T_N$ is extremely weak above 11 GPa and the superconducting $T_c^{onset}$ reaches 6.5 K at 13.3 GPa. These results further confirm that our findings are reliable and reproducible.

Based on the above high-pressure results, we construct a temperature-pressure phase diagram of KMn$_6$Bi$_5$ single crystal as shown in the Fig. 4(a). With increasing pressure, $T_N$ first increases slightly till 4.0 GPa, then shows a sudden decrease at 6.0 GPa, and collapses completely at $P_c \approx 13$ GPa. The pressure dependence of $T_N(P)$ at $P \geq 6.2$ GPa can be fitted by the empirical power model $T_N = T_0(1-P/P_c)^\beta$, where $\beta$ represents the critical exponent. The best fitting yields $T_0 = 70.7$ K, $P_c = 13.1$ GPa and $\beta = 0.25$ for $KMn_6Bi_5$, respectively. Bulk superconductivity emerges at about 12.5 GPa, coexists with antiferromagnetic order in a narrow pressure range of 12.5-12.9 GPa, and increases continuously after the collapse of antiferromagnetic order above $P_c$. But, the slope of $dT_c^{onset}/dP$ is reduced gradually with pressure and displays a superconducting dome on the border of long-range antiferromagnetic order, which is similar to many unconventional superconducting systems [1-6]. In concomitant enhancement of $T_c(P)$ from 12.9 to 14.2 GPa, the $\mu_0 H_{c2}(0)$ also increases dramatically with increasing pressure as shown in Fig. 4(b).

The close proximity of superconductivity to a magnetic instability suggests that the critical magnetic fluctuations may play an important role in mediate Cooper pairing. To substantiate this point, we analyzed the normal-state resistivity just above $T_c$, which can be fitted by the power law, i.e. $\rho = \rho_0 + A \times T^n$, where $\rho_0$ represents the residual resistivity, the coefficient $A$ and the exponent $n$ are associated with the density of states at Fermi level and the inelastic electron scattering, respectively. As shown in Fig. 4(c), the exponent $n$ initially decreases from $\sim 2.0$ for Fermi liquid at 12.5 GPa to a minimum $\sim 1.0$ for no-Fermi liquid at 14.2 GPa, implying the breakdown of Fermi-liquid state around $P_c$; the reduction of $n$ is closely related with the collapse of antiferromagnetic order that produces enhanced critical spin fluctuations. Although the enhancement of coefficient $A$ is not significant as typically seen near a magnetic QCP, the broad maximum of $A$ is positively correlated with the evolution of $T_c(P)$. Thus, the observations of non-Fermi-liquid behavior and moderate enhancement of coefficient $A$ near $P_c$ in $KMn_6Bi_5$ signal the important role of magnetic fluctuations in the transport properties [4-7].

As mentioned above, the $T$-$P$ phase diagram of $KMn_6Bi_5$ featured by the emergence of superconductivity on the border of antiferromagnetic order resembles those of many unconventional superconductors associated with the magnetism-medicated Cooper pairing mechanism. Its unconventional nature is also substantiated by the observations of the strange-metal behavior in the normal-state resistivity near $P_c$ and an unusually larger $\mu_0 H_{c2}(0)$ exceeding the Pauli limit. Previous theoretical calculations in the sister compound $RbMn_6Bi_5$ suggest that the density of states at the Fermi level is dominated by the Mn-3d orbital electrons at AP. If the electronic structures are retained under pressure, $KMn_6Bi_5$ becomes the first ternary Mn-based superconductor with a relatively high $T_c$ of 9.3 K, which is one order of magnitude higher than that of MnP (~ 1 K) [7].

The discovery of superconductivity in the ternary $KMn_6Bi_5$ is important for the following reasons: (i) the intriguing crystal structure renders a complex magnetic structure within the inner Mn-cluster column, making it a good candidate to study the geometrically frustrated magnetism and unconventional Cooper pairing; (ii) it proves that the transition temperature of Mn-based superconductors can be raised to the level of 10 K; (iii) the Q1D magnetic materials with enhanced magnetic fluctuations can serve as a new paradigm for finding more Mn-based superconductors, as exemplified also in the cases of $A_2Cr_3As_3$ [14-17] and $ACr_3As_3$ [18,19] ($A$ = K, Rb, Cs or Na); (iv) in comparison with the binary MnP, the crystal structure and chemical compositions of ternary $AMn_6Bi_5$ are more flexible and offer more possibilities to explore. For example, one can tune the magnetic and transport properties at AP through chemical substitutions at either $A$- or Bi-sites so as to induce superconductivity, e.g. alkaline-earth (Ca, Sr, Ba)-doped $AMn_6Bi_5$, or replacing Bi with Sb/Te in $AMn_6Bi_5$. On the other hand, whether the infinite $[Mn_6Bi_5]^-$ columns could sever as a new superconducting gene deserves further investigations. Electronic structure and superconducting properties of $AMn_6Bi_5$ upon varying the $A$-site cations also deserve in-depth investigations under pressure.

## Conclusion

In summary, we discovered the first ternary Mn-based superconductor $KMn_6Bi_5$ under high pressure. We found that bulk superconductivity emerges in the pressurized single-crystal $KMn_6Bi_5$ when its antiferromagnetic order is suppressed by pressure, resulting in a superconducting dome in the vicinity of magnetic order. The close proximity of superconductivity to a magnetic instability suggests an unconventional paring mechanism. The optimal $T_c^{onset}$ reaches ~ 9.3 K at 14.2 GPa, and the zero-temperature upper critical field $\mu_0H_{c2}(0)$ is found to exceed the Pauli limit. Our present study opens a new avenue for finding more Mn-based superconductors.

## Acknowledgements

This work is supported by the Beijing Natural Science Foundation (Z190008), National Key R&D Program of China (2018YFA0305700), the National Natural Science Foundation of China (Grant Nos. 12025408, 11874417, 11874400, 11834016, 11921004, 11888101), the Strategic Priority Research Program and Key Research Program of Frontier Sciences of CAS (XDB25000000, XDB33000000 and QYZDB-

**Figure caption:**

**FIG. 1.** (a) Crystal structure of single-crystal $KMn_6Bi_5$. (b) Temperature dependence of resistivity $\rho(T)$ and its derivative $d\rho/dT$ along the $b$-axis at ambient pressure (left axis). The antiferromagnetic transition at $T_N$ is marked by vertical dotted line. Temperature dependence of magnetic susceptibility $\chi(T)$ (right axis) recorded at a magnetic field of 0.2 T. The Curie-Weiss fitting is shown by the broken line.

**FIG. 2.** (a) Electrical resistance of $KMn_6Bi_5$ (#1) under various hydrostatic pressures up to 14.2 GPa measured in CAC. (b) The corresponding $dR/dT$ curves are shifted vertically for clarity. Inset of (a) shows the low-$T$ resistance data to highlight the evolution of the antiferromagnetic transition.

**FIG. 3.** (a) The enlarged view of low-$T$ resistance data ($0 \leq T \leq 11$ K), in which the $T_c^{onset}$ and $T_c^{offset}$ are marked by arrows. (b) The ac magnetic susceptibility $4\pi\chi_{ac}(T)$ of $KMn_6Bi_5$ (#2) and Pb with similar volume. The arrows show the superconducting transition temperature $T_c^{\chi}$ of $KMn_6Bi_5$ crystal. (c) The low-temperature $R(T)$ under different fields at 14.2 GPa. (d) The temperature dependences of $\mu_0H_{c2}(T)$ fitted by the Ginzburg-Landau (GL) equation; the Pauli limit of $\mu_0H_p = 1.84T_c$ at 14.2 GPa is indicate.

**FIG. 4.** (a) Temperature-pressure phase diagrams for $KMn_6Bi_5$. The characteristic temperatures for the antiferromagnetic (AFM) transition $T_N$ and the superconducting (SC) transition $T_c^{onset}$, $T_c^{offset}$ and $T_c^{\chi}$ are determined from the resistance and ac magnetic susceptibility measurement on samples #1 to #4. Pressure dependences of (b) $\mu_0H_{c2}(0)$ and (c) the coefficient $A$ (left) and the exponent $n$ (right).

## Figure 1

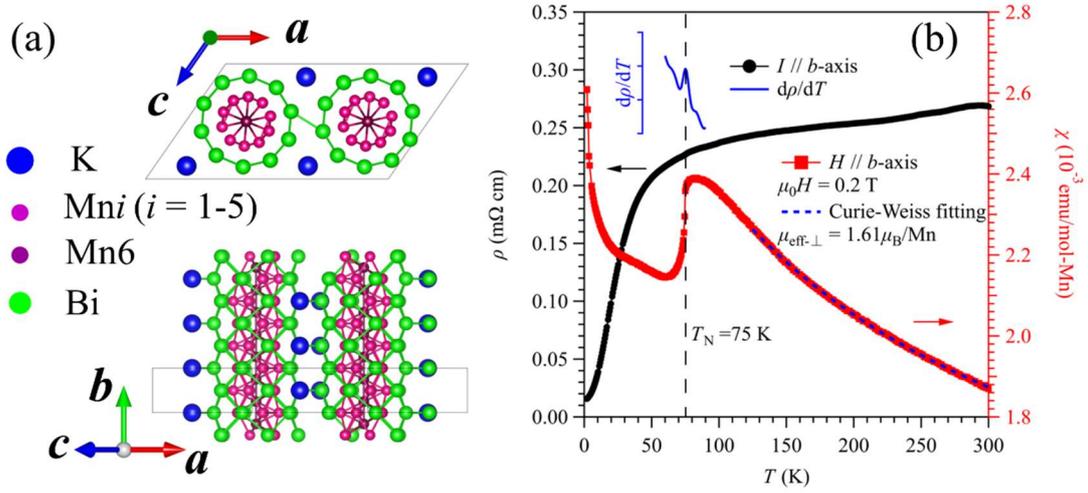

## Figure 2

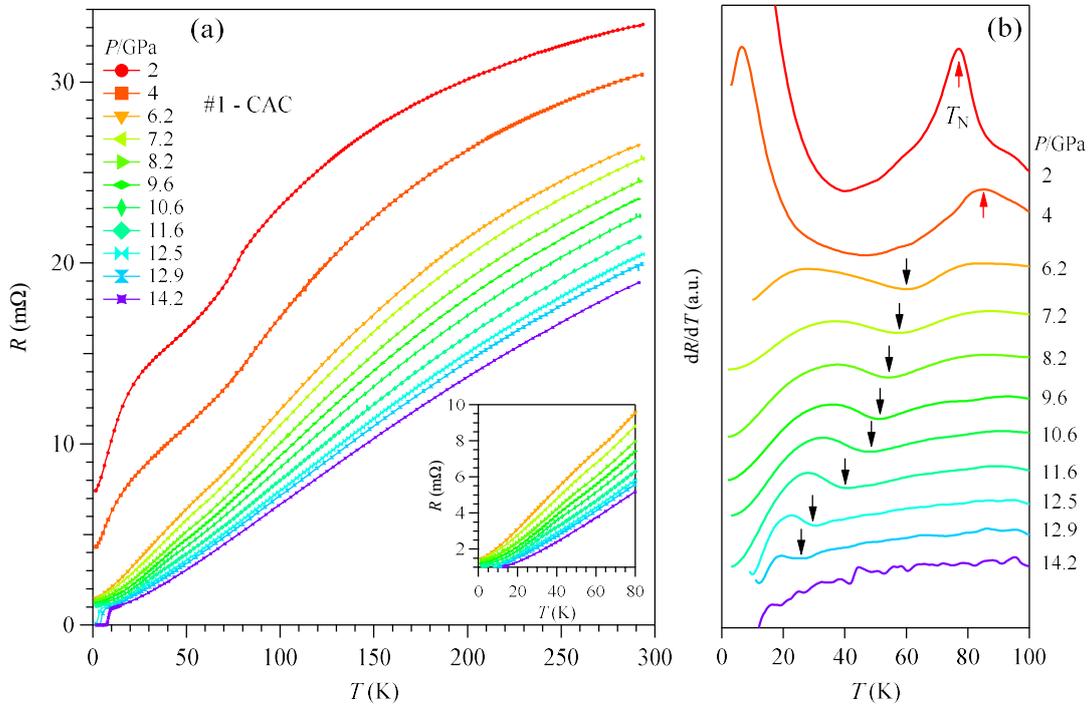

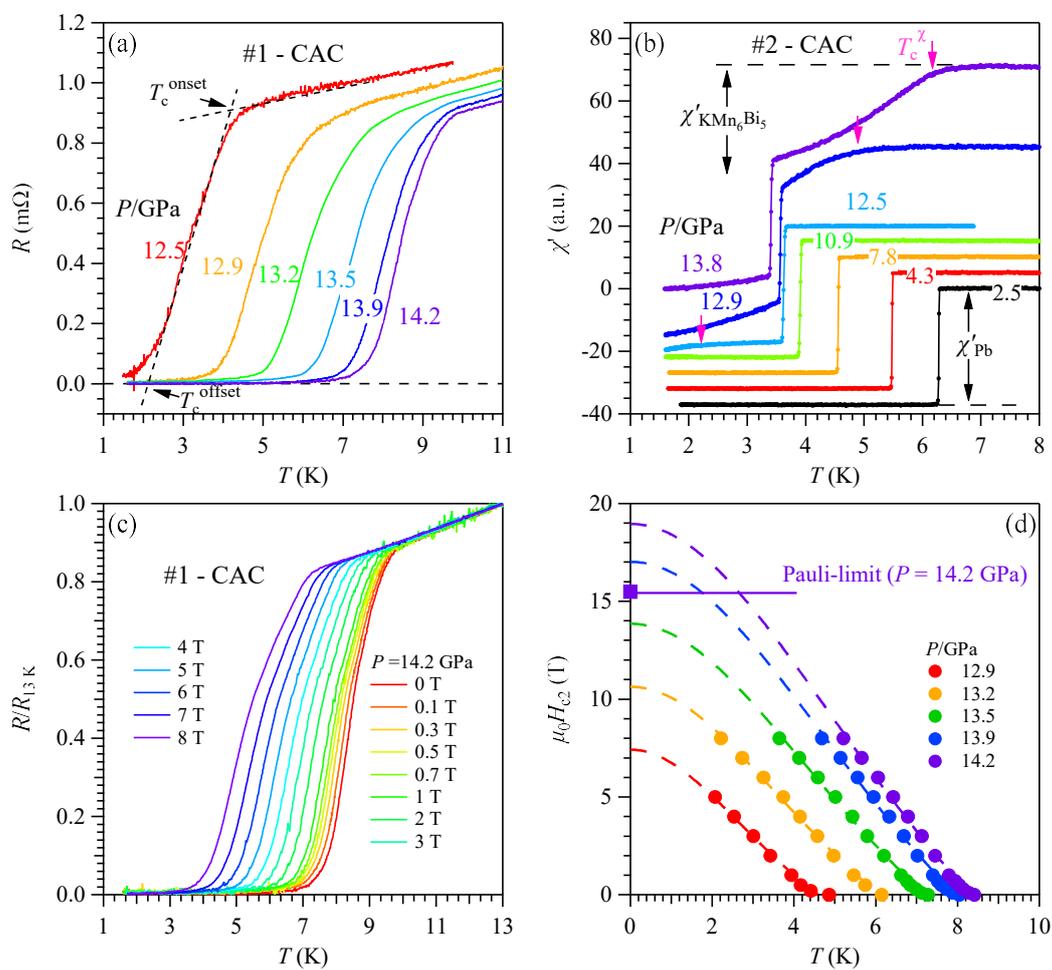

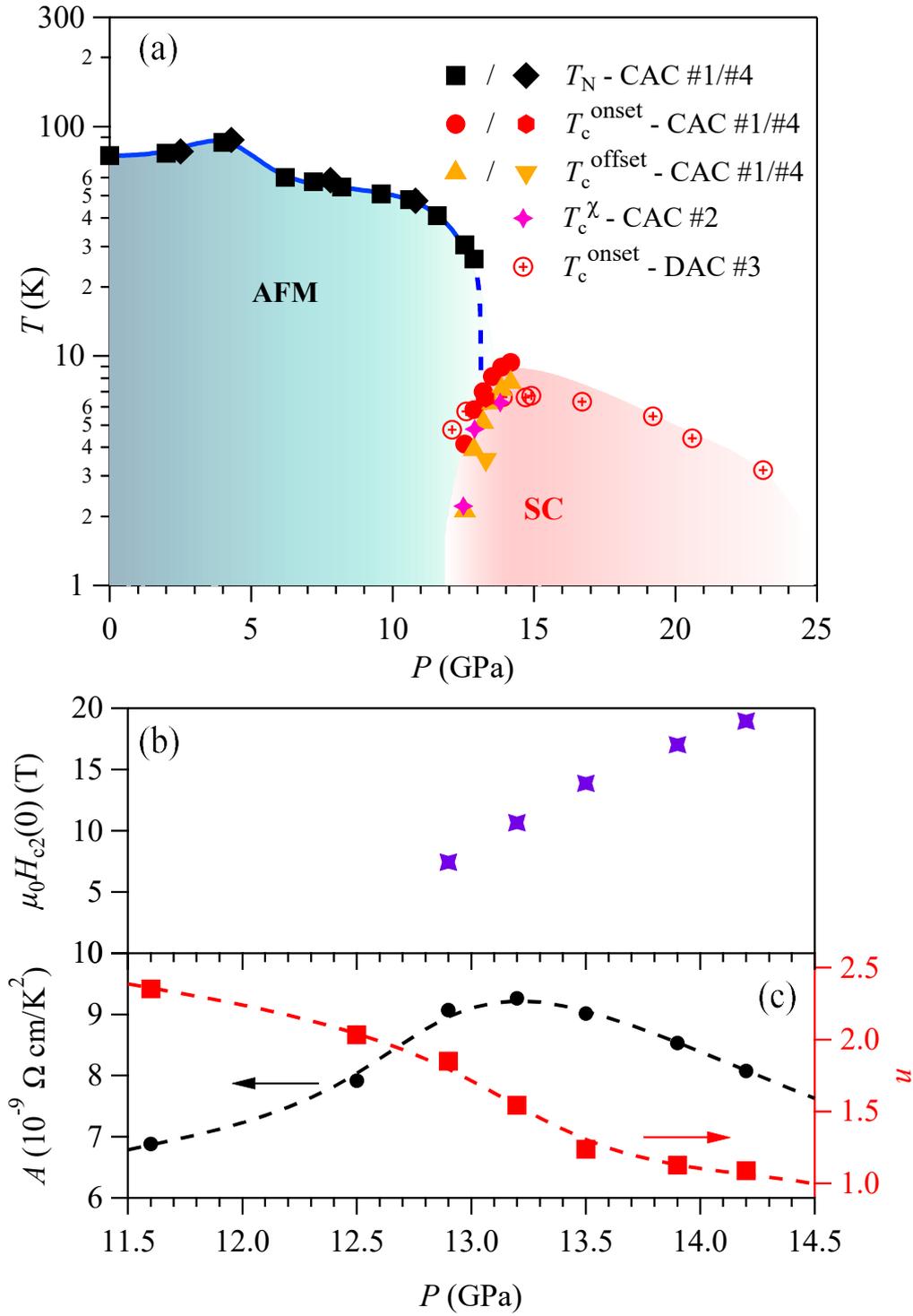